\documentstyle[12pt]{article}  
\begin{document}
\pagestyle{plain}\textheight=21truecm\textwidth=13truecm
\begin{center}
\begin{bf}
VORTICES IN PROTOPLANETARY DISKS 
\end{bf}  
\end{center}
\vspace{1.cm}  
\begin{center}
Patrick Godon \& Mario Livio \\
\vspace{5.cm}  
Space Telescope Science Institute \\
3700 San Martin Drive \\ 
Baltimore, MD 21218 \\
\vspace{0.5cm}  
e-mail: godon@stsci.edu, mlivio@stsci.edu  \\ 
\end{center}
\newpage

\baselineskip 15pt plus 2pt
\begin{center}
\bf{Abstract}
\end{center}

We use a high order accuracy spectral code 
to carry out two-dimensional time-dependent numerical simulations
of vortices in accretion disks. 
In particular, we examine the stability and the life time of 
vortices in circumstellar disks around young stellar objects. 
The results show that cyclonic vortices dissipate quickly, 
while anticyclonic vortices can survive in the flow for hundreds of
orbits.
When more than one vortex is present,
the anticyclonic vortices interact through vorticity waves 
and merge together to form larger vortices. 
The exponential decay time $\tau$ of anticyclonic vortices is 
of the order of $30-60$ orbital periods for a viscosity parameter 
$\alpha \approx 10^{-4}$ 
(and it increases to $\tau \approx 315$ for $\alpha = 10^{-5}$), which   
is sufficiently long to allow heavy dust particles to rapidly
concentrate in the core of anticyclonic vortices in protoplanetary disks. 
This dust concentration increases the local density of centimeter-size grains,
thereby favoring the formation of larger scale objects which are then 
capable of efficiently triggering a gravitational instability. 
The relatively long-lived   
vortices found in this work may therefore play a key role in the 
formation process of giant planets. \\  

Subject Headings: accretion, accretion disks - circumstellar matter - 
hydrodynamics - planetary systems - stars: formation -
stars: pre-main sequence    

\newpage 
\section{Introduction}

It is presently believed that the formation of the planets in the 
solar system took 
place via a progressive aggregation of dust grains in the 
primordial solar nebula.  
However, a mechanism for building planetesimals
between the centimer-size grains (formed by agglomeration
and sticking) and the meter-size objects capable of triggering
a gravitational instability is still lacking 
(see e.g. Adams \& Lin 1993 and the recent review of Beckwith,
Henning \& Nakagawa 1999).   
Recently, it has been suggested 
(Barge \& Sommeria 1995; Adams \& Watkins 1995; Tanga et al. 1996), 
that heavy dust particles rapidly concentrate in the cores
of anticyclonic vortices in the solar nebula, increasing the local
density of centimeter-size grains and favoring the formation of larger
scale objects which are then capable of efficiently triggering a gravitational
instability. The gravitational instability itself is also enhanced
because a vortex creates a larger local surface density in the disk. 
Consequently, even if giant planets form (as an alternative model
suggests) initially due to a gravitational instabilty of the gaseous
disk (rather than by planetesimal agglomeration), vortices may still
play an important role. 
It is the change in the Keplerian velocity of the flow in the disk,
due to the anticyclonic motion in the 
vortex, that induces a net force towards the 
center of the vortex (the inverse happens in a cyclonic vortex). 
As a consequence, 
within a few revolutions of the vortex around the protostar, 
the concentration of dust grains 
and the density in the anticyclonic vortex 
become much larger than outside.  
Analytical estimates (Tanga et al. 1996) have shown that,  
depending on the unknown drag on the dust particles in the gas, the 
trapping time for dust in a small anticyclonic vortex (of size $D/r 
\approx 0.01$ or less) located at 5 AU 
(the distance of Jupiter from the Sun) is about
$20 < \tau < 10^3 $years, i.e. between about 2 and 100 local 
periods of revolution. The mass that can accumulate in the core of
the vortex on this time scale is of the order of $10^{-5}$ earth
masses (i.e. a planetesimal).  
Barge \& Sommeria (1995) considered larger
vortices (of size $D \approx H$, where $H$ is the disk half
thickness, and $H/r \approx 0.06$ at 5AU) 
and obtained that on a timescale of the order of 500
orbits, a core mass of about 16 earth masses can accumulate in the
core of the vortex. Since small vortices are expected to merge
into larger ones, these results should be regarded as complementary. 
The results of
Tanga et al. (1996) could represent the early stages in planet formation 
(planetesimals), while the results of Barge \& Sommeria 
(1995) could represent a later stage in the evolution (the formation of 
planets cores).    \\ 
 
It is therefore important for the theory of planet formation
to assess under which conditions vortices
form in disks, whether they are stable or not, and for how long they
can survive in the disk. \\  

Analytical results (Adams \& Watkins 1995, using a geostrophic approximation) 
have shown that many different
types of vortex solutions are possible in circumstellar disks. However, 
Adams \& Watkins (1995) used simplifying assumptions and they were unable to 
estimate the lifetime and stability of the vortices. 
In order to resolve the issue of the non-linear behavior of vortices
(e.g. vortex merging, scattering, growth, etc.) one needs to carry out 
detailed numerical simulations of the flow. \\   
 
recent simplified simulations of a disk (Bracco et al. 1998),    
which assume an incompressible flow and solve the shallow water
equations, indicate that 
two-dimensional (large scale) vortices do not fragment into small 
vortices because of the inverse cascade of energy (characteristic of 
two-dimensional flows). Rather, the opposite happens:
small vortices merge together to form and sustain a large vortex.
It is believed for example that the 
Great Red Spot in the atmosphere of Jupiter
is a steady solitary vortex also sustained by the merging of small
vortices (Marcus 1993; see also Ingersoll 1990). However, in this case
the strong winds in Jupiter's atmosphere are also partially feeding
vorticity from the background flow into the vortex, and the decay time
of the vortex is much longer.  \\ 
 
The shear in Keplerian disks inhibits the formation of vortices larger than a 
given scale length $L_s$ ($L_s^2=v/(\partial \Omega / \partial r )$, where
$v$ is the rotational velocity of the vortex (assumed to be subsonic, otherwise
the energy is dissipated due to sound waves and shocks), and
$\Omega$ is the angular velocity
in the disk). At the same time, 
the viscosity dissipates quickly vortices smaller than a viscous  
scale length $L_{\nu}$.  
Consequently, only vortices of size $L$ satisfying $L_{\nu} < L < L_s$ 
can survive in the flow for many orbits before being dissipated. 
In the calculations of Bracco et al. (1998), 
positive density perturbation counter rotating (anticyclonic) vortices 
were found to be 
long lived, while cyclonic vortices dissipated very quickly. 
However, as we noted above, Bracco et al. (1998) assumed  
a shallow water approximation 
for the disk. 
These authors do not specify 
the value of the specific heats ratio  
$\gamma$ that was used in the simulations 
(in order for the shallow water equation to represent a polytropic
thin disk one has to assume $\gamma = 2$; see e.g. Nauta \& T\'oth 1998).
Most importantly the Mach number of the Keplerian flow 
(or equivalently the thickness of the disk) is not defined.
It is presently not clear whether  
incompressible results (using a shallow water approximation) are
valid for more realistic disks, i.e. for compressible (and potentially
viscous) disks, with a density
profile $\rho \propto r^{-15/8}$ and a temperature profile $T\propto
r^{-3/4}$ - the standard Shakura-Sunyaev disk model (Shakura \& Sunyaev
1973).     \\

It is the goal of the present work to model vortices in disks, using
a more realistic approach. We  
conduct a numerical study of the stability and lifetime of vortices 
in a standard disk model of a protoplanetary nebula. For this purpose,  
we use a time-dependent, two-dimensional high-order accuracy 
hydrodynamical compressible code, assuming a polytropic 
relation. Here it is important to stress the following.  
The vorticity (circulation) of the flow (with a velocity field 
$\vec{u}$) is defined as 
\begin{equation}
\vec{w} = \vec{ \nabla } \times \vec{u} .
\end{equation}
Taking the curl of the Navier-Stokes Equations 
(see e.g. Tassoul 1978) one obtains an 
equation for the vorticity 
\begin{equation}
\frac{\partial \vec{w}}{\partial t} + \vec{u} . \vec{\nabla}
\rho \vec{w} \propto \vec{\nabla} P \times \vec{\nabla} \rho + ... 
\end{equation} 
The first term on the right hand side of the equation is the source
term for the vorticity. 
Therefore, vorticity can be generated by the non-alignment
of $\nabla \rho$ with $\nabla P$ (e.g. meridional circulation
in stars is generated by this term). The resultant growth of 
circulation due to this term is also known as the 
baroclinic instability. 
However, in the present work we do not address the question 
of the vortex generation process, since we are using a polytropic
relation $P=K \rho ^{\gamma}$ ($K$ is the polytropic
constant, $\gamma = 1 + 1/n$, and $n$ is the polytropic index),
and the source term vanishes. In this case the vortencity
(vorticity per unit mass) is conserved (Kelvin's circulation theorem,
see e.g. Pedlosky 1987) and vortices cannot be generated. \\ 

In the next section we present the numerical modelling of the protoplanetary
disk. The results are presented \S 3 and a discussion follows. \\  


\section{Accretion Disks modelling} 

The time-dependent 
equations (e.g. Tassoul 1978) were written in cylindrical
coordinates $(r, \phi, z)$, and were further
integrated in the vertical (i.e. $z$) direction (e.g. Pringle 1981). 
We use a Fourier-Chebyshev spectral method 
of collocation (described in Godon 1997)  
to solve the equations of the disk.  
Spectral Methods are global high-order accuracy methods  
(Gottlieb \& Orszag 1977; Voigt, 
Gottlieb \& Hussaini 1984; Canuto et al. 1988). 
These methods are fast and accurate 
and are frequently used to 
study turbulent flows (e.g. She, Jackson \& Orszag 1991)
and interactions of vortices (Cho \& Polvani 1996a). 
It is important to stress that spectral codes have very little
numerical dissipation, and that the only dissipation that occurs
in the present simulations is due to the $\alpha$ viscosity introduced
in the equations.  
All the details of the numerical method and the exact form of the 
equations can be found in Godon (1997). We therefore give here only 
a brief overview of the modelling.  \\  

The equations
are solved in the plane of the disk $(r, \phi )$, with
$0 \le \phi \le 2 \pi$ and $R_0 \le r \le 2 R_0$, where
the inner radius of the computational domain, $R_0$, has been
normalized to unity. 
We use an alpha prescription (Shakura \& Sunyaev 1973)
for the viscosity law, 
$ \nu = \alpha c_s H = \alpha c_s^2 / \Omega_K $, 
where $c_s$ is the sound speed and $H = c_s/\Omega_K$ 
is the vertical height of the disk.
The unperturbed flow is Keplerian, and  
we assume a polytropic relation $P=K\rho^{1+1/n}$ ($n$ 
is the polytropic index, while the polytropic constant $K$ is fixed 
by choosing $H/r$ at the outer radial boundary). 
Both the inner and the outer radial boundaries are treated as 
free boundaries, i.e. with non-reflective boundary conditions
where the conditions are imposed on the characteristics of the flow
at the boundaries.  \\  

The Reynolds number in the flow is given by 
\[ R_e = \frac{L u}{\nu}, \] 
where $L$ is a characteristic dimension of the computational domain,
$u$ is the velocity of the flow (or more precisely, the velocity 
change in the flow over a distance $L$) 
and $\nu$ is the viscosity.
Since we are solving for the entire disk, $L \approx r$ and 
$u \approx v_K$, and the Reynolds number becomes
\[ R_e = \alpha^{-1} (H/r)^{-2}, \]
where we have substituted $\nu = \alpha H^2 \Omega_K$, since 
we are using an $\alpha$ viscosity prescription. The Reynolds number
in the flow is very high (of the order of $10^4-10^5$ for the assumed
parameters). \\  

The simulations were carried out without the use of spectral filters and with
a moderate resolution of $128 \times 128$ collocations points. As we noted 
above, the only 
dissipation in the flow is due to the $\alpha$ viscosity introduced in the
equations (i.e. the Navier-Stokes equations). \\ 

\section{Numerical Results} 

In all the models presented here we chose $H/r=0.15$ to match protoplanetary disks,
however similar results were obtained using $H/r=0.05$ and $H/r=0.25$. 
The models were first evolved in the radial dimension for an initial
dynamical relaxation phase (lasting several Keplerian orbits at least).
Then an axisymmetric disk was constructed 
from the one-dimensional results, on 
top of which we introduced an initial vorticity perturbation.  
The initial vorticity perturbation had a Gaussian-like form and a 
constant rotational velocity of the order of $\approx 0.2 c_s$.  
In all the models we found that cylonic vortices dissipate very quickly,
within about one local Keplerian orbit, 
while anticyclonic disturbances persist in the flow for a much
longer period of time. 
It is important to stress that in all the models, the 
anticyclonic vortex, although rotating 
in the retrograde direction (in the rotating frame of reference), 
has a rotation rate that is slower than the 
local Keplerian flow, and consequently, in the inertial frame of 
reference the vortex rotates in the prograde direction, like a planet.    
As the models are evolved, 
the initial vorticity perturbation is streched and 
elongated by the shear, within a few Keplerian orbits.  
The elongation of the vorticity into a thin structure transfers 
enstrophy (the potential enstrophy is defined as the average of the
square of the potential vorticity) 
towards high wave numbers. This process is consistent with
the direct cascade of enstrophy characteristic of two-dimensional
turbulence.  
It forms an elongated negative (relative to the local flow) vorticity strip
in the direction of the shear, 
with an elongated vortex in the middle of it (Figure 1). Due to a
Kelvin-Helmoltz instability, perturbations in 
a forming vorticity strip propagate (along the strip) 
in the direction opposite to the
shear. These propagating waves are called Rossby waves in Geophysics 
(see e.g. Hoskins et al. 1985; in Geophysics this instability is referred to
as a shearing instability, e.g. Haynes 1987; Marcus, 1993, section
6.2). 
As a result of this instability the two bendings in the vorticity strip 
(namely, the trailing and the leading vorticity waves of the vortex) 
move in the direction opposite to the shear and fold onto themselves. 
The extremities of the vorticity waves can be again elongated by the shear
and and can undergo further folding due to the propagation of the Rossby waves. 
This results in spiral vorticity arms around the vortex (Figure 2).    
The shape of the vortex thus formed, then does not change any more 
during the remaining time of the simulation (dozens of orbits). \\  

We also find that anticyclonic vortices act
like overdense regions in the disk, i.e. within about one orbit the density 
increases by about 10 percent in the core of the vortex. We cannot 
check, however, whether a cyclonic vortex decreases the density in its
core, since cyclonic vortices dissipate very quickly. \\

\subsection{Flat density profile models} 

In a first series of models, we chose a constant density profile
throughout the disk with a polytropic index $n=3$. These models are
less realistic and are probably closer to the models of Bracco et al.
(1998) who used an incompressible approximation (the shallow water
equation).  
We ran four models with different values of the viscosity parameter
$\alpha = 1 \times 10^{-4}$, $3 \times 10^{-4}$, $6 \times 10^{-4}$ 
and $1 \times 10^{-3}$. The viscosity parameter was chosen so as to
be consistent with values inferred for protostellar disks
(e.g. Bell et al. 1995). 
The simulations were followed for up to a maximum time of 
60 local Keplerian orbits of the vortex in the disk. 
We found exponential decay times for the vortex of
$\tau = 3.9$ periods for $\alpha=10^{-3}$ and  
$\tau = 32.4$ periods for $\alpha=10^{-4}$ (see Figure 4).
In all cases the decay was exponential.

\subsection{Standard disk models} 

In an attempt to study the stability and decay times of vortices
in more realistic disks, we modelled a standard Shakura Sunyaev disk 
(Shakura \& Sunyaev, 1973) with 
$\rho \propto r^{-15/8}$ and $T \propto r^{-3/4}$ (this was achieved
by choosing a polytropic index $n=2.5$ together with an ideal
gas equation of state). In this model we chose $\alpha = 10^{-4}$, 
and the simulations were followed for up to 20 local Keplerian orbits 
of the vortex in the disk. Although the initial vorticity perturbation
was the same as before, the decay time of the vortex ($\tau = 60$),
and its size were found to be larger than for the flat density models 
(see Figure 4).  
However, difficulties in assessing precisely the size of the vortex
make it difficult to determine whether the different decay time is due
to the different size of the vortex or merely to the different density
profiles of the models. We also ran an additional model with 
a lower viscosity parameter of $\alpha = 10^{-5}$ and found 
an exponential decay time of more than 315 orbits. In this case the
resolution was increased to $256 \times 256$ and the vortex that formed was
slightly smaller in size than in the previous models.  \\  

In order to gain further insight into the dynamics of vortices,
we also ran
two additional models (one with $\alpha=10^{-4}$ and 
one with $\alpha = 10^{-3}$) where two vorticity perturbations
were initially introduced in the flow. In the case of $\alpha = 
10^{-4}$, the vortices interact by propagating vorticity
waves and eventually the two vortices merge together to form a single
larger vortex (see Figures 5a - 5d). 
In the case of $\alpha=10^{-3}$, the vortices do not
interact (no vorticity waves were observed) and dissipate quickly.  \\  

The results indicate that the exponential decay time of a vortex
is inversely proportional to the viscosity (Figure 6), and it can be
very large indeed 
(in $\approx 30-60$ orbits the amplitude of the vortex decreases by
a factor of $e$) for disks around Young Stellar Objects, where
the viscosity might be low ($\alpha = 10^{-4}$). One expects the decay time
to behave like $\tau \propto d^2/\nu$ (see also Bracco et al. 1998), 
where $d$ is the size of the vortex
and $\nu$ is the viscosity in the flow.
In principle, an inviscid model 
could have vortices that do not decay. Therefore, any decay that has
previously been observed in inviscid numerical simulations (e.g. the 
shallow water approximation of Bracco et al. 1998) was probably due to 
numerical diffusion in the code (the hyperviscosity introduced
by Bracco et al. 1998 in their model). \\

\section{Discussion}

Accretion disks possess a very strong shear, which is normally 
believed to lead to 
a rapid destruction of any structures that form within it 
(e.g. the spiral 
waves obtained in a perturbed disk by Godon \& Livio 1998,
dissipate or exit the computational domain within about 10 orbits
for $\alpha = 1 \times 10^{-4}$). However, we find that anticyclonic vortices 
are surprisingly 
long-lived, and they can survive for hundreds of orbits (the amplitude
of the vortex decreases exponentially with a time constant of 60 orbits
for $\alpha = 10^{-4}$ in disks around young stellar objects).     
These results are in agreement with other similar simulations of
the decay of two-dimensional turbulence, e.g. the simulations of
rotating shallow-water decaying turbulence on the surface of a sphere
(Cho \& Polvani, 1996a \& b; modelling of the Jovian atmosphere), 
where the only dissipation of the vorticity is due to the (hyper) viscosity.  
The size and the elongation of the vortices that form out 
of an initial vorticity perturation increase with the viscosity. 
In order for the model to be self consistent,
the size of the vortices has to be larger than the thickness of the disk
(validity of the two-dimensional assumption, otherwise the vortices
are three dimensional). We found that for an alpha viscosity of the order of
$10^{-5}$ (with $H/r=0.15$) the semi-major axis $a$ of the vortices 
is slightly smaller than $H$. However, when the viscosity is increased 
to $10^{-3}$ the semi-major axis becomes up to three 
times the thickness of the disk. In all the cases the 
semi-minor axis $b$ remains smaller than $H$, while the elongation
($a/b$) varies from about 4 (for the less viscous case) to about 10 (for the
most viscous model).   
Tanga et al. (1996) and Barge \& Sommeria (1995) solved numerically the  
equations of motion for dust particles in vortices,  
and confirmed that dust particles concentrate inside vortices on a relatively
short timescale. The time taken by a dust particle to reach the center of 
an anticyclonic vortex at a few AU ranges from a few orbits
to a hundred orbits, depending on the exact value of the 
drag parameter. We have shown that in a standard disk model  
for a protoplanetary disk (a polytropic disk with $H/r=0.15$), 
vortices can survive long enough to allow dust particles to 
concentrate in their core. 
For $\alpha=10^{-3}$ only small vortices would form and would not
merge together. In this case (using the estimate of Tanga et al.
1996 for small vortices) one finds that the vortices would decay
within about 10 orbits and the dust concentration in the core
of the vortices would only reach $10^{-6}$ Earth masses (planetesimals, 
at 5AU).
For $\alpha =10^{-4}$ small vortices would merge together to
form larger vortices. The concentration of dust grains in the core
of the larger vortices could reach about 2 Earth masses (within
a hundred revolutions and using the estimate of Barge \& Sommeria 1995
at 5AU). 
Therefore, the local density of centimeter-size grains
could be increased, thus 
favoring the formation of larger scale objects which are then 
capable of efficiently triggering a gravitational instability. 
Our results therefore confirm earlier suspicions that were based on 
a simplified solution of shallow water equations for an incompressible fluid. \\

It is important to note though, that in the present work we have not
addressed yet the problem of vortex formation. In addition to the baroclinic
instability described in the Introduction, another potential way to 
generate vortices in disks around young stellar objects
is through infall of rotating clumps
of gas. It has been suggested that protostellar disks could grow
from the accretion (or collapse) of rotating gas cloud (e.g. Cassen
\& Moosman 1981; Boss \& Graham 1993; Graham 1994; Fiebig 1997). 
The clumps with the proper rotation vectors could
then give rise to small vortices, that would subsequently 
merge together. \\ 

Finally, we would like to mention that vortices can 
(in principle at least) have many other
astrophysical applications. For example, interacting Rossby waves 
can result in radial angular momentum transport 
(e.g. Llewellyin Smith 1996).  Vortices
are also believed to be important in molecular cloud substructure formation
in the Galactic disk (e.g. Chantry, Grappin \& L\'eorat 1993; Sasao 1973).\\  

\section*{Acknowledgments} 
This work has been supported in part by NASA Grant NAG5-6857 and
by the Director Discretionary Research Fund at STScI. \\  

\newpage 
\section*{References}

\noindent       
Adams, F. C., \& Lin, D. N. C. 1993, in Protostars and Planets III,
ed. E. H. Levy \& J. I. Lunine (Tuscon: Univ.Arizona Press), 721 
\\ \\          
Adams, F. C., \& Watkins, R. 1995, ApJ, 451, 314 
\\ \\          
Barge, P., \& Sommeria, J. 1995, A \& A, 295, L1 
\\ \\
Beckwith, S. V. W., Henning, T., Nakagawa, Y., 1999, in Protostars
and Planets IV, in press, astro-ph/9902241  
\\ \\   
Bell, K. R., Lin, D. N. C., Hartmann, L. W., \& Kenyon, S. J.,
1995, ApJ, 444, 376 
\\ \\ 
Boss, A. P., Graham, J. A., 1993, ICARUS, 106, 168 
\\ \\ 
Bracco, A., Provenzale, A., Spiegel, E., Yecko, P. 1998, in Abramowicz
A. (ed.), Proceedings of the Conference on Quasars and Accretion Disks,
Cambridge Univ. Press, astro-ph/9802298  
\\ \\          
Canuto, C., Hussaini, M. Y., Quarteroni, A. \& Zang, T. A. 1988, 
Spectral Methods in Fluid Dynamics (New York: Springer Verlag) 
\\ \\          
Cassen, P., \& Moosman, A. 1981, ICARUS, 48, 353 
\\ \\          
Chantry, P., Grappin, R., \& L\'eorat, J. 1993, A \& A, 272, 555 
\\ \\          
Cho, J. Y. K., \& Polvani, L. M. 1996a, Phys. Fluids, 8 (6), 1531 
\\ \\          
Cho, J. Y. K., \& Polvani, L. M. 1996b, Science, 273, 335  
\\ \\          
Fiebig, D., 1997, A \& A, 327, 758 
\\ \\ 
Godon, P., \& Livio, M. 1998, submitted to ApJ 
\\ \\ 
Godon, P. 1997, ApJ, 480, 329 
\\ \\          
Gottlieb, D., \& Orszag, S.A. 1977, Numerical Analysis of Spectral
Methods: Theory and Applications (NSF-CBMS Monograph 26;
Philadelphia: SIAM) 
\\ \\ 
Graham, J. A., 1994, AASM, 184, 44.07 
\\ \\ 
Haynes, P.H., 1987, J.Fluid Mech., 175, 463 
\\ \\ 
Hoskins, B., McIntyre, M., Robertson, A., 1985, Q.J.R.Meteorol.Soc., 111, 877 
\\ \\ 
Ingersoll, A. P., 1990, Science, 248, 308 
\\ \\ 
Marcus, P. S. 1993, ARAA, 31, 523 
\\ \\       
Nauta, M. D., \& T\'oth, G. 1998, A \& A, 336, 791 
\\ \\       
 Pedlosky, J. 1987, Geophysical Fluid Dynamics, 2nd ed.
 (Springer Verlag, New York)
 \\ \\ 
Shakura, N.I., \& Sunyaev, R. A. 1973, A \& A, 24, 337. 
\\ \\       
She, Z.S., Jackson, E., \& Orszag, S.A. 1991, Proceedings of the Royal
Society of London, Series A: Mathematical and Physical Sciences,
vol. 434 (1890), 101. 
\\ \\       
Sasao, T. 1973, PASJ, 25, 1 
\\ \\       
Tanga, P., Babiano, A., Dubrulle, B., \& Provenzale, A., 1996, ICARUS, 121, 158 
\\ \\       
Tassoul, J. L., 1978, Theory of Rotating Stars, Princeton Series in
Astrophysics, Princeton, New Jersey 
\\ \\       
Voigt, R.G., Gottlieb, D., \& Hussaini, M.Y. 1984, Spectral Methods for
Partial Differential Equations (Philadelphia: SIAM-CBMS) 
\\ \\      
%

\newpage 

{\bf Figures Caption}  

{\small{\it Figure 1:
A colorscale of the vorticity shows the stretching of an 
anticyclonic vorticity perturbation  
in a disk model with a viscosity parameter of $\alpha = 10^{-4}$.
This results in a negative vorticity strip in the flow, which is then
further affected by the propagation of Rossby waves along its wings.  
}} \\  

{\small{\it Figure 2:
A detailed view of an anticyclonic vortex in the disk, a few orbits later 
than shown in Figure 1.
The vorticity wings have folded up onto themselves due to a Kelvin-Helmoltz
instability (accompanied by the propagation of Rossby waves).    
}} \\  

{\small{\it Figure 3: 
The momentum ($\rho \vec{v}$) field is shown in the vicinity of the vortex. The
coordinates ($r,\phi$) have been displayed on a Cartesian grid.  
}} \\  

{\small{\it Figure 4: The maximum amplitude $A$ of the absolute vorticity
of the vortex is drawn in arbitrary logarithmic units as a function
of time (in orbits), for values of the alpha viscosity parameter ranging
from $\alpha = 10^{-5}$ to $\alpha = 10^{-3}$. The full lines  
represents the initial flat density profile models, while the doted lines
are for an initial density profile matching a standard disk model.
In all the models, during the first orbits, 
the decay of the maximum amplitude of the vorticity
is stronger, due to
the relaxation of the initial vorticity perturbation. 
The figure shows that the amplitude behaves
like $A \propto e^{-t/\tau}$, where $\tau$, the decay time,
increases as the viscosity decreases. 
}} \\  

{\small{\it Figures 5a-d. As two vortices interact, they emit vorticity
waves and eventually merge to form one large vortex.
In this model the viscosity parameter was taken to be 
$\alpha = 10^{-4}$. For convenience
the vortices are shown roughly in the same orientation in the disk. 
The complete process of merging takes about $5-10$ orbits.}}   
\\
 
{\small{\it Figure 6: The decay time $\tau$ against $\alpha^{-1}$.  
The flat-density models of Figure 4 are represented by stars, together with
a straight line $\tau = c/\alpha$, where $c=3.2 \times 10^{-3}$.  
}} \\

\end{document}